# Micropores can enhance intrinsic fracture energy of hydrogels

Puyu Cao[1], Bin Chen[1,2*], Yi Cao[3,4], Huajian Gao[5]

[1]Department of Engineering Mechanics, Zhejiang University, Hangzhou, People's Republic of China

[2]Key Laboratory of Soft Machines and Smart Devices of Zhejiang Province, Hangzhou, People's Republic of China

[3]School of Physics, Nanjing University, Nanjing, China

[4]Collaborative Innovation Center of Atmospheric Environment and Equipment Technology, Jiangsu Key Laboratory of Atmospheric Environment Monitoring and Pollution Control, School of Environmental Science and Engineering, Nanjing University of Information Science & Technology, Nanjing 210044, P. R. China

[5]Mechano-X Institute, Applied Mechanics Laboratory, Department of Engineering Mechanics, Tsinghua University, Beijing 100084, China

*To whom correspondence should be addressed: chenb6@zju.edu.cn.

**Abstract**

It is widely known that hydrogels, a class of soft materials made of a polymer chain network, are prone to fatigue failure. To understand the underlying mechanism, here we simulate polymer scission and fatigue initiation in the vicinity of a crack tip in a two-dimensional chain network. For a network without pores, our findings reveal that polymer scission can take place across multiple layers of chains, rather than just a single layer as assumed in the classical Lake-Thomas theory, in consistency with previus studies. For a network with a high density of micropores, our results demonstrate that the pores can substantially enhance the intrinsic fracture energy of the network in direct proportion to the pore size. The underlying mechanism is attributed to pore-pore interactions which lead to a relatively uniform distribution of cohesive energy ahead of the crack tip. Our model suggests that micropores could be a promising strategy for improving the intrinsic fracture energy of hydrogels and that natural porous tissues may have evolved for enhanced fatigue resistance.

**Introduction**

In recent years, synthetic hydrogels have undergone significant advancements and are increasingly being utilized across various fields such as tissue engineering, agriculture, drug delivery, soft robotics, and flexible devices, among others (1–4). Many of these applications require hydrogels to withstand cyclic loads, including repetitive loadings on hydrogel-based biomimetic skins (5) and cartilage (6), even though hydrogels are known to be susceptible to fatigue failure (7). To combat this challenge, a number of hydrogels with elevated fatigue thresholds have been developed following various strategies (8–10).

The fatigue threshold of hydrogels is mainly attributed to their intrinsic fracture energy, which has traditionally been evaluated (11–16) using the Lake-Thomas theory (17), focusing on the breakage of a single layer of polymer chains across the crack plane (Fig. 1a) without triggering inelastic deformation in the bulk. The thickness of a chain layer in its unstrained state can be approximated, for instance, by the equilibrium separation between two ends of an individual polymer chain in the freely-jointed chain model. Theoretical predictions have demonstrated good agreement with experimental findings across various polymeric materials, such as PAAm-Ca-alginate (18, 19), PAAm-PAMPS (12), and PAAm-polyvinyl alcohol (13). Nevertheless, a source of uncertainty stems from estimating the thickness of the energy-dissipating layer, and open questions remain regarding why chain scission appears confined to a single layer rather than involving multiple layers of chains (18).

High-density pores are often introduced in hydrogels through phase separation during synthesis (20). SEM images of hydrogels reveal a multitude of micropores with diameters of ~2 μm (Fig. 1b), through which microbeads can diffuse around (Fig. 1c). The average pore size (21), pore size distribution, and interconnections between pores are crucial aspects of a hydrogel matrix. Notably, in freeze-dried hydrogels derived from silk fibroin solutions, pore sizes diminish with increasing silk fibroin concentration and decrease with rising temperature at a constant silk fibroin concentration (22). Hydrogels freeze-dried with $Ca^{2+}$ exhibit larger pore sizes compared to those without $Ca^{2+}$ ions in the silk fibroin aqueous solutions (23). Various technologies, including solvent casting particle leaching (24), freeze-drying (25), gas foaming (26), and electrospinning (27), have been deployed to regulate the microarchitectural features of pores in hydrogels and facilitate the creation of functional hydrogels that mimic native tissue structures facilitating cell viability (28), proliferation (29) and migration (30).

This study aims to investigate the effect of micropores on the intrinsic fracture energy of the polymer network in a hydrogel. We begin by validating the Lake-Thomas theory (17) using a two dimensional (2D) chain network model with a pre-existing crack. Subsequently, we simulate the polymer chain scission and fatigue initiation in the vicinity of the crack in a 2D network containing a high density of micropores. Our findings reveal that the pores result in altered stress distribution in the material, leading to a significant enhancement of the intrinsic fracture energy that can scale with the pore size. These findings highlight the potential for designing hydrogels with specific microarchitectural pore features to improve the intrinsic fracture energy.

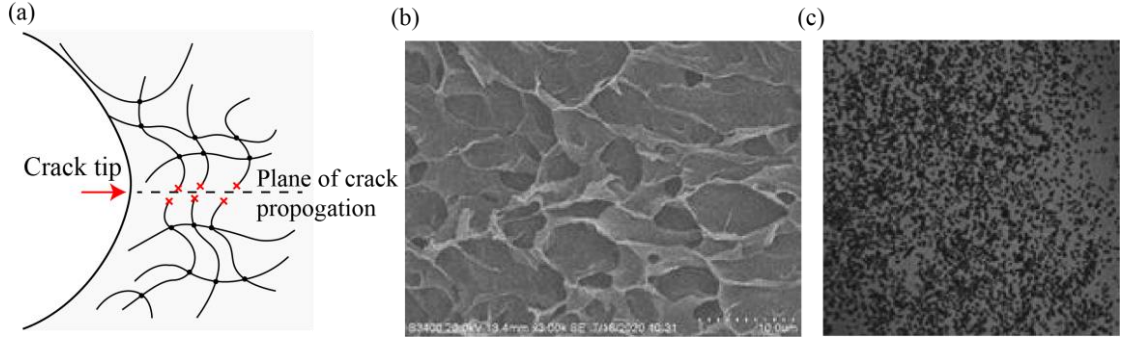

Fig. 1. Classical Lake-Thomas theory and evidence for the existence of micropores in hydrogels. (a) A schematic illustration of the Lake-Thomas theory which assumes that the intrinsic fracture energy of a polymer is due to the scission of a layer of polymer chains along the prospective crack plane; (b) An SEM image showing small pores existing in a hydrogel through which microbeads of a diameter of 2 μm are observed to undergo diffusive migration (c).

**Validation of the Lake-Thomas theory**

Several polymer chain network models were recently developed (31–34) to investigate the mechanical behaviors of polymeric materials. To validate the classical Lake-Thomas theory, we have constructed a 2D polymer chain network in COMSOL, which is convenient to use and different from most of the previous polymer chain network models. Our chain network is composed of truss elements representing individual polymer chains as well as solid elements for sustaining 2D hydrostatic pressure. Within the network, polymer chains are periodically distributed and form triangular meshes. A prestress is introduced in the network (e.g., due to water swelling) such that each polymer chain of initial length $l_0$ displays the same end-to-end distance of $l_d$. Individual nodes of truss elements physically correspond to crosslinks in the chain network. The worm-like chain model (35) is chosen to describe the force-stretch behavior of each individual polymer chain, $F = \frac{kT}{P}\left(\frac{1}{4}\left(1-\frac{x}{L_c}\right)^{-2} - \frac{1}{4} + \frac{x}{L_c}\right)$, where $k$ is Boltzmann's constant, $T$ the temperature, $P$ the persistence length, $L_c$ the contour length, and $x$ the length of the polymer chain. 2D solid elements are adopted to maintain the in-plane near incompressibility of a neo-Hookean hyperelastic material, with a sufficiently large bulk modulus $\kappa$ and a small shear modulus $G$, ensuring an approximately equibiaxial (2D hydrostatic) stress state.

To use the polymer chain network model to simulate the intrinsic fracture energy, we consider a 2D strip with a crack under uniaxial loading, as illustrated in Fig. 2a. In the simulation, only half of the polymer chain network is considered, with symmetry boundary conditions applied at the bottom edge along the crack propagation path as well as the left edge which is sufficiently far away from the crack tip, for better convergence.

Typical force contours and deformation from the simulation are displayed in Fig. 2b. The simulation is terminated when the force of any individual chain reaches a set critical value $F_c$, referred to as the chain breaking force. The intrinsic fracture energy of the network is then evaluated through the J-integral (36),

$$J = \varepsilon_0 h, \qquad (1)$$

where $\varepsilon_0$ is the strain energy density far ahead of the crack tip and $h$ the height of the strip. As displayed in Fig. 2c, the simulated intrinsic fracture energy increases with $F_c$.

According to the Lake-Thomas theory (17), the intrinsic fracture energy $\Gamma_0$ could be expressed as $\Gamma_0 \simeq w_o \bar{L}$, where $\bar{L}$ is the initial chain length typically ~10 nm, and $w_0$ the critical strain energy density for chain rupture at the crack tip. Following the force-stretch curve of a worm-like chain, $w_0$ can be calculated for given $F_c$, and the prediction according to the Lake-Thomas theory is plotted in Figs. 2c,d.

As shown in Figs. 2c, d, our results agree well with the prediction by the Lake-Thomas theory at relatively low $F_c$ or $w_0$, where the corresponding bond breaking force falls below ~100 pN. At relatively high $F_c$ or $w_0$, e.g., as the bond breaking force reaches ~1 nN, our simulated intrinsic fracture energy is significantly larger than the prediction of the Lake-Thomas theory, with ~500% discrepancy. To understand the origin of such discrepancy, we plot the force contour of individual chains in the vicinity of the crack tip. It can be seen from Figs. 3a-c that the chain forces are highly concentrated at the crack tip and decay rapidly over a few neighboring chains. The decaying distance along the vertical direction increases with $F_c$. We let the crack tip advance by breaking one single polymer chain at the crack tip and calculate the associated changes in chain force within the network. The results in Figs. 3d-f reveal that the dissipation zone associated with crack propagation can involve a few layers in the vertical direction and the larger the $F_c$, the more layers are involved.

The above results indicate that multiple layers of chains are involved in crack advance at fatigue threshold, thereby challenging the assumption in the Lake-Thomas theory that only a single layer of polymer chains contributes to the intrinsic fracture energy of a polymeric material, especially when the chain breaking force can reach that of a covalent bond on the order of ~1 nN (37). Nevertheless, our simulation results are in agreement with the commonly held view in the literature (7) that the intrinsic fracture energy of a polymer network, within which the force-stretch curve of individual chain obeys the worm-like chain theory (35) and the chain breaking force is below ~2 nN, generally scales with the initial length of individual chains, i.e.

$$J_0 \sim w_0 \bar{L}. \tag{2}$$

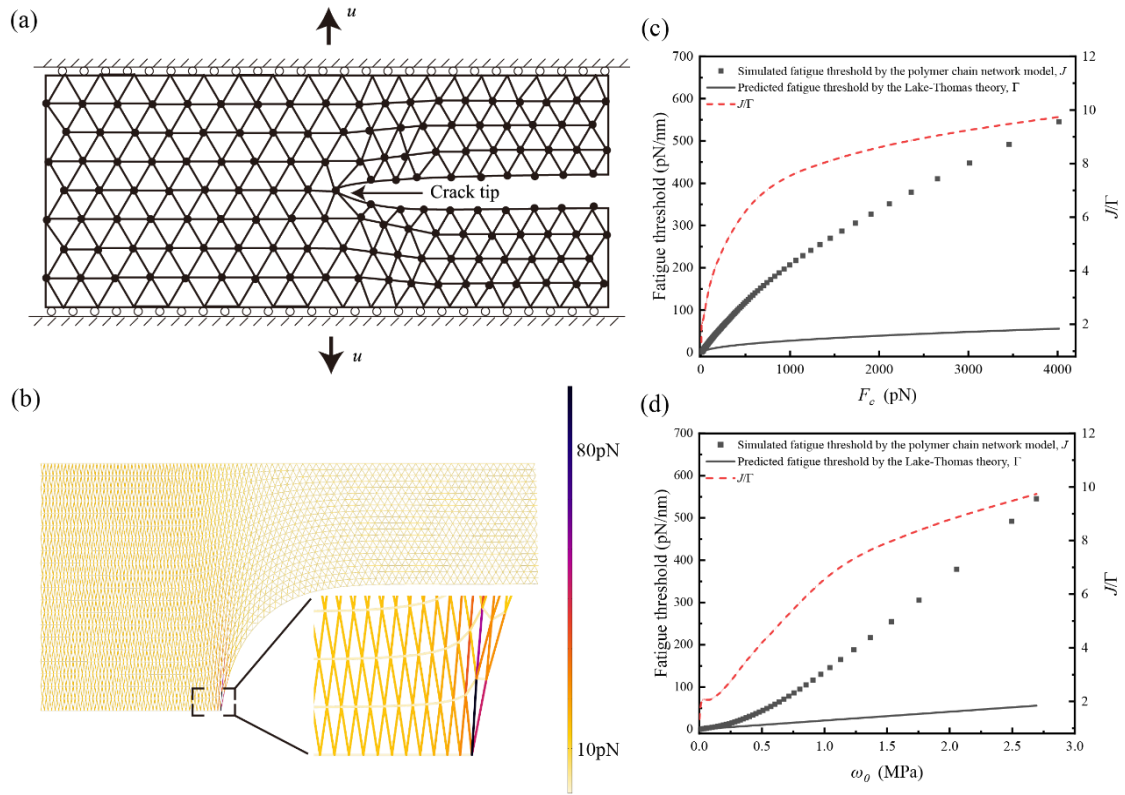

Fig. 2. Comparison between simulated intrinsic fracture energy and prediction from the Lake-Thomas theory. (a) Simulation of a crack propagating in a strip of triangular polymer chain network. (b) Simulated chain force contours and deformed configuration. Comparison between the simulated intrinsic fracture energy with prediction from the Lake-Thomas theory as functions of (c) critical chain force and of (d) critical strain energy density.

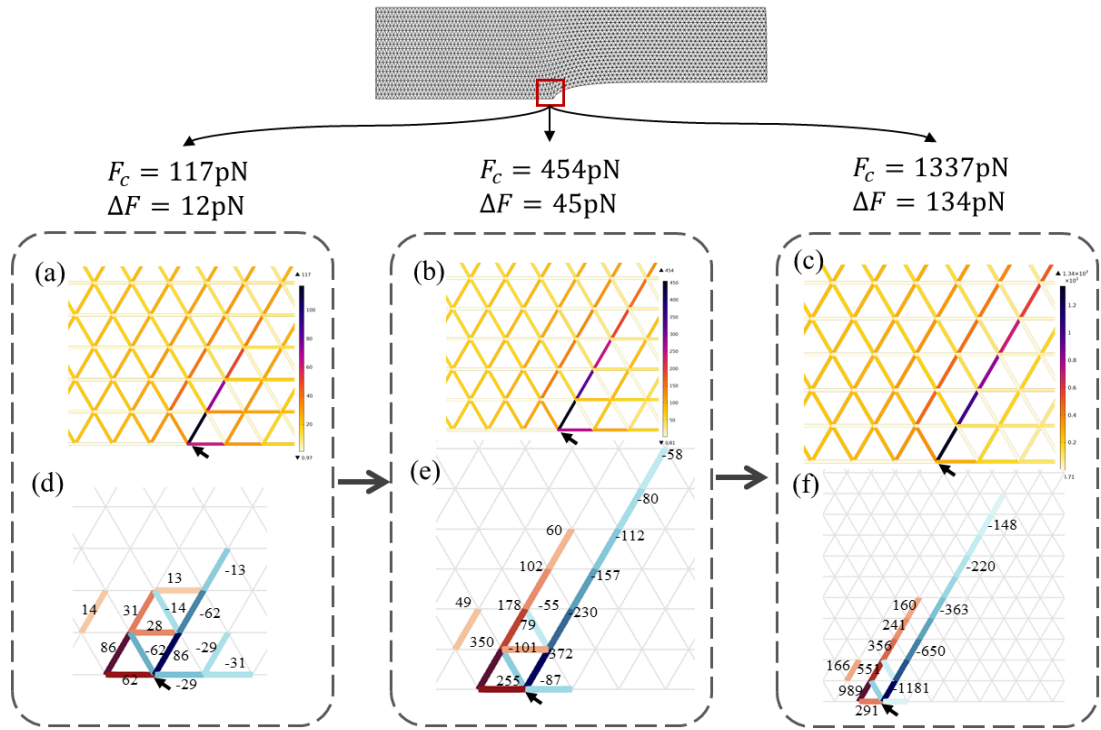

Fig. 3. Simulated crack tip force and deformation fields. (a-c) Chain force contours in the vicinity of the crack tip under different $F_c$. (d-f) Change in chain forces upon crack propagation by breaking a single polymer chain at the crack tip with different $F_c$. Only the change in chain force above a small threshold value, $\Delta F$, is plotted, with $\Delta F = 12$ pN in (d), 45 pN in (e) and 134 pN in (f), respectively.

# Result

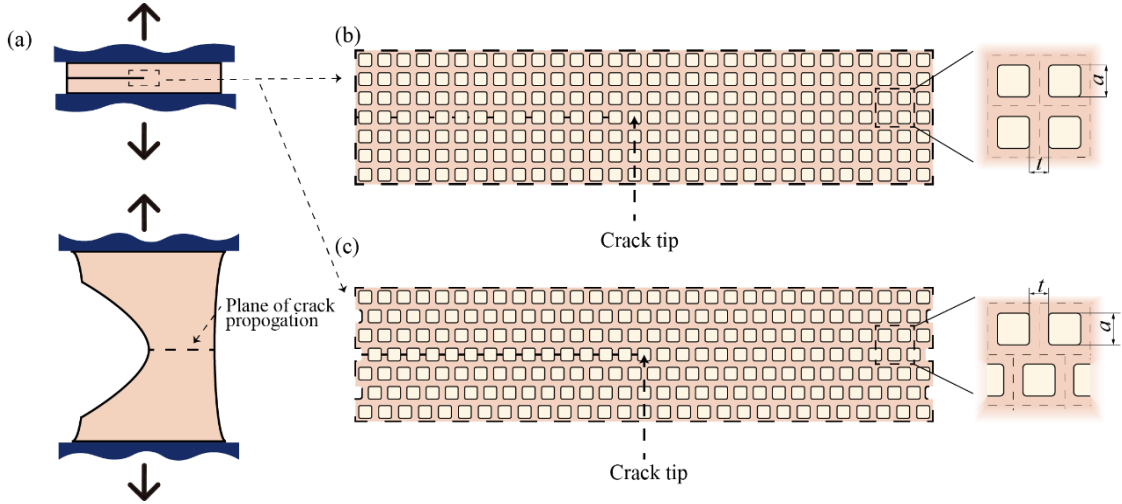

Fig. 4 Simulation of porous polymer networks. (a) Mode I crack propagation in a strip of hydrogel with (b) stacked and (c) staggered pores.

Next we turn to investigate the intrinsic fracture energy of the network with periodically distributed pores. Similar to experiments (38), the pure-shear model of a plane stress cracked strip is employed for determining the intrinsic fracture energy (Fig. 4a), where the pores of radius $a$ are taken to be square shaped with rounded corners and uniformly distributed in two patterns, the stacked and the staggered, as illustrated in Figs. 4b,c. The walls between neighboring pores have consistent thickness $t$. Since a soft or debonded inclusion tends to attract cracks (39), we forcus on the scenario where the crack tip is situated within a pore.

Using a two-scale approach, we integrate the polymer network model with the traditional finite element method (FEM) for simulating a strip containing a high density of pores. This two-step methodology begins with an initial simulation using FEM in COMSOL. Taking advantage of symmetry, only half of the strip is modeled. The strip length is chosen to be sufficiently large and the horizontal displacement is constrained at the far right side of the strip to furtherly aid convergence. For simplicity and without loss of generality, in the FEM we employ the incompressible Arruda-Boyce model (40) which incorporates two material parameters: the locking stretch $\lambda_m$ of polymer chains, and the shear modulus $\mu$.

At the lower scale, our attention shifts toward a localized region near the crack tip, where the displacements calculated from FEM are imposed as boundary conditions to model the forces and stretches within the chain network. For the truss elements, the polynominal series of the Langevin chain model (41–43) is chosen to describe the force-stretch behavior of individual chains, $F = \frac{kT}{2P}\left(3\frac{x}{Lc} + \frac{9}{5}\left(\frac{x}{Lc}\right)^3 + \frac{297}{175}\left(\frac{x}{Lc}\right)^5 + \frac{1539}{875}\left(\frac{x}{Lc}\right)^7 + \frac{126117}{37375}\left(\frac{x}{Lc}\right)^9\right)$. In addition, solid elements are adopted to model a nearly incompressible neo-Hookean hyperelastic materials with a large bulk modulus of $\kappa$ and a small shear modulus of $G$. The hydrodatic stress in the solid elements is enforced as $F_3/\left(\frac{\sqrt{3}}{2}l_d^2\right)$, where $F_3$ is the Langevin chain force with extension $\lambda_3 l_d$, $\lambda_3$ being the out-of-plane stretch. Note that, in the Arruda-Boyce model, $\mu$ is obtained by fitting the stress-stretch curves under uniaxial tension and $\lambda_m = \sqrt{L_C/2P}/\lambda_s$, $\lambda_S$ denoting the the free-swelling stretch.

The two-scale simulation proceeds until the maximum chain force in the vicinity of the crack tip reaches a critical threshold value, $F_c$, at which point we evaluate the average strain energy $w_1$ far ahead of the crack tip and determine the intrinsic fracture energy $J$ as the product of $w_1$ and the strip height (36) under two sets of parameters: The first set varies the pore size under fixed wall thickness, while the second set changes the wall thickness at fixed pore size. The computed $J/w_0$ are displayed in Fig. 5. The simulation results shown in Fig. 5 indicate that $J/w_0$ can significantly exceed $\bar{L}$ for both pore patterns. The simulated $J$ increases almost linearly with pore size under fixed wall thickness (Fig. 5a). Conversely, enlarging the wall thickness under fixed constant pore size only affects $J/\omega_0$ slightly (Fig. 5b). The effect of the chain breaking force, $F_c$, is displayed in Fig. 5c, where $J/(w_0 a)$ increases with $F_c$.

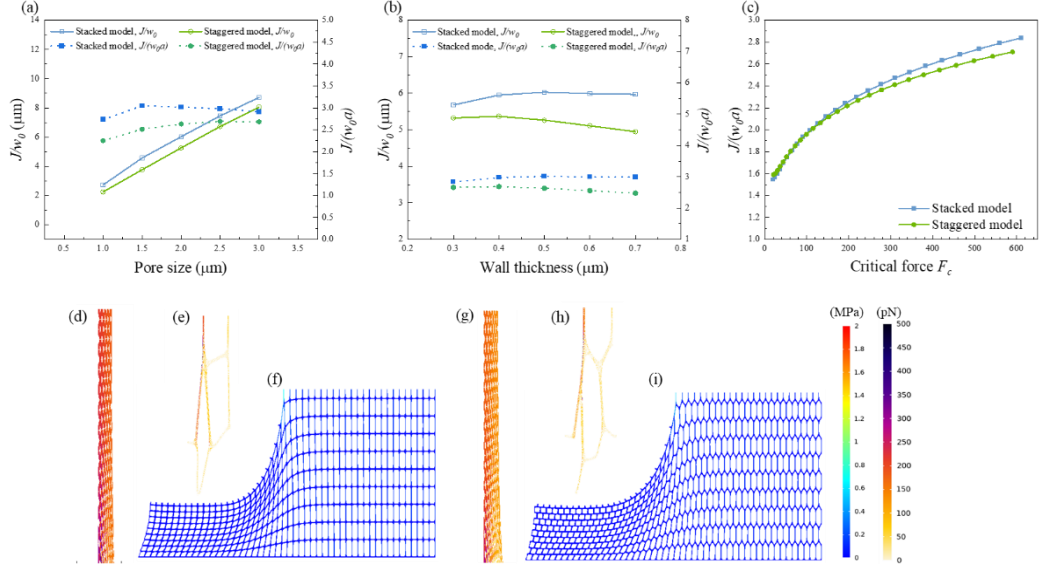

Fig. 5. (a) Simulated $J/w_0$ is nearly proportional to the pore size under fixed wall thickness, and the nondimensional value $J/w_0 a \sim 2.5$ for all cases; (b) Under fixed pore size, $J/w_0$ varies little with the wall thickness and $J/w_0 a \sim 2.5$ for all cases; (c) $J/w_0 a$ increases with $F_c$; Typical simulated force contours within the walls at the crack tip in the stacked model (d, e) and in the staggered model (g, h); Typical simulated strain energy density contours in the stacked model (f) and staggered model (i). Note that the strain energy density distribution is almost uniform at to the crack tip.

To make sense of the simulation results, we examine the deformation and force distribution in the crack tip region. As shown in Figs. 5d-i for both stacked and staggered pore patterns, the walls directly at the crack tip are largely under uniaxial loading with nearly uniform strain energy density distribution. To understand why the intrinsic fracture energy scales with the pore size, one can note that the walls between pores near the crack tip can be regarded as a type of super cohesive bonds, where the stored elastic energy would be released upon crack propagation (Fig. 6). For an interface with periodic cohesive energy distribution, it has been shown that the apparent fracture energy falls between the peak and the average values of the cohesive energy, depending on the relationship between the size of the cohesive zone, $l_c$, and the period of the cohesive energy distribution, $l_p$ (44). For $l_c \ll l_p$, the apparent fracture energy of the interface is the peak value of the cohesive energy. On the other hand, when $l_p \ll l_c$, the apparent fracture energy of the interface becomes equal to the average value of the cohesive energy.

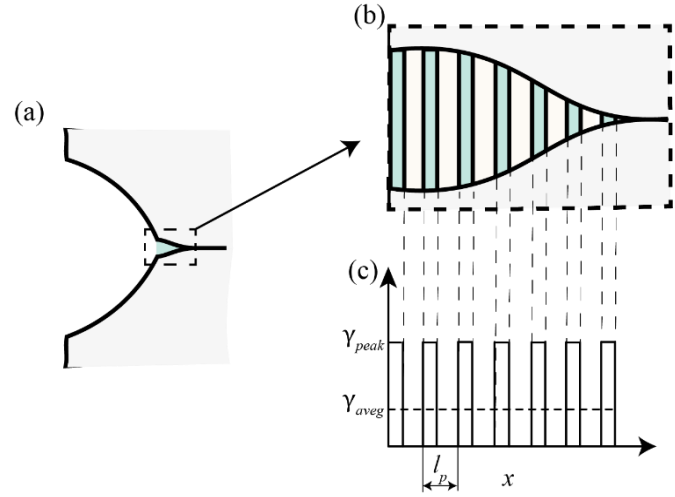

Fig. 6. A cohesive model of crack (a) with periodic super cohesive bonds (b) representing the walls between pores in a porous hydrogel network. The periodic cohesive bonds can be modeled as a periodic distribution of cohesive energy (c).

In the case of a highly porous hydrogel network displayed in Fig. 4, the average cohesive energy is

$$\gamma_{\text{aveg}} \sim \frac{w_0 t \cdot a}{a+t} \sim w_0 t, \tag{3}$$

which roughly scales with the wall thickness between neighboring pores, considering that $a$ is generally larger than $t$. The peak value of cohesive energy is

$$\gamma_{\text{peak}} \sim \frac{w_0 t \cdot a}{t} = w_0 a, \tag{4}$$

which scales with the pore size.

As shown in Figs. 5a,b, the simulated intrinsic fracture energy is roughly

$$J_1 \sim 2.5\, w_0 a, \tag{5}$$

suggesting that the intrinsic fracture energy corresponds more to the peak value of cohesive energy in Eq. (5) in our current analysis. This is consistent with the theory of Chen et al. (2008) (45) when the cohesive zone size at the onset of fatigue is relatively small compared to the pore size. The numerical factor of 2.5 from simulation indicates that energy stored in multiple layers of walls could contribute to the intrinsic fracture energy.

**Discussion and conclusion**

A chain network model has been developed to investigate the intrinsic fracture energy of hydrogels and compared to the Lake-Thomas theory which relates the intrinsic fracture energy to the breaking of a single layer of polymer chains along the crack plane. With force-stretch curves of individual chains within the network obeying the worm-like chain theory (35), our findings generally align with the Lake-Thomas theory in that the intrinsic fracture energy scales with the initial length of the constituent polymer chain, with a chain breaking force that is not excessively high. Importantly, our analysis demonstrates that chain scission is usually not confined to one layer of chains but can instead involve multiple layers of chains, as recently speculated in the literature (18).

Our anlayis is in consistency with a most recent study (46), where the energy dissipation from chains far from the crack tip was found to contribute to the intrinsic fracture energy of polymer networks. In that work (46), the intrinsic fracture energy of polymer networks were also found to be several times of the Lake-Thomas prediction, when the ratio of the stiff energetic modulus to the soft entropic modulus of indivdula chains modified in their polymer network model was ~1000 (46). However, when this ratio increased to a very high value, for example, $2 \times 10^4$, together with the chosen of a very high chain breaking force of 5 nN, the intrinsic frauctre energy was found to be almost 2 orders of magnitude higher than the Lake-Thomas prediction (46). Please note that incompressibility was not enforced in their polymer newework model (46), which should be generally obeyed in soft polymeric materials and has been realized in our polymer network model.

Our analysis further indicates that the intrinsic fracture energy of a highly porous network can scale with the pore size, instead of the chain length. Considering that the chain length is typically on the order of ~10nm and pore size on the order of ~1μm, this means that the intrinsic fracture energy of a highly porous network could reach orders of magnitude higher than that of the homogenous network. In other words, our analysis indicates that the micropores with soft materials could highly magnify the intrinsic fracture energy.

The intrinsic fracture energy of polymer networks, as measured in experiments (11, 16, 47), were found to be nearly two orders of magnitude greater (46) than the Lake-Thomas prediction. This substaintial discrepancy was previously attributed to the nonlocal energy dissipation mechanism involving the relaxation of chains far from the crack tip, which was closely tied to the high ratio of the stiff energetic modulus to the soft entropic modulus of individual chains within the polymers (46). However, based on Eqs. (2-4), our studies here suggest an alternative explanation: the presence of micorpores within fabricated polymers may be a contributing factor to this substantial difference in the intrinsic fracture energy.

In our simulations, the distribution of strain energy density across the wall thickness between neighbouring pores is nearly uniform. However, such uniformity will not hold as the wall thickness becomes sufficiently large and this could lower the amount of strain energy that can be stored prior to the fatigue threshold, leading to a intrinsic fracture energy that is lower than the one predicted above. In this sense, it is important to note that there exists a critical wall thickness for flaw tolerance, below which the strain energy density distribution could stay uniform (48). Thus introducing sufficiently small, dense, hierarchically structured pores with sufficiently thin walls could be a viable method to achieve relatively uniform distribution of strain energy density and enhance the intrinsic fracture energy.

Indeed, the fabrication of hydrogels with hierarchical pores appears to be a promising area. For instance, the integration of directional freeze-casting and subsequent salting-out treatments has recently enabled the fabrication of hydrogels with multi-length-scale hierarchical structures (49). These hydrogels boast micrometre-scale honeycomb-like pore walls, which consist of interconnected nanofibril meshes. Despite a water content of up to 95 per cent, these hydrogels demonstrate high strength, toughness, fatigue resistance, etc., making them comparable to other robust hydrogels and even natural tendons.

Our analysis also suggests that the ubiquitous pores within native tissues may play a role in their exceptional fatigue properties. For instance, both experimental and simulation data have shown that the presence of pores in echinoderm stereo can reduce stress concentration within the stereom, enabling high relative strength and significant energy absorption capabilities (50). Similarly, diffused damage in bone, often resulting from daily activities, is typically ~1 μm in length and has been shown to be highly effective in energy absorption (51). Interestingly, age-related reductions in the formation of such diffused damage might partly explain the deterioration of fracture properties in bones with age.

## Acknowledgements

This work was supported by Zhejiang Provincial Natural Science Foundation of China (Grant No.: LZ23A020004) and the National Natural Science Foundation of China (Grant No.: 11872334)

**Table I** Default parameters used in the worm-like chain model.

| $l_d$ | $P$ | $L_c$ | $kT$ | $G$ | $\kappa$ |
|---|---|---|---|---|---|
| 24 nm | 0.4 nm | 80 nm | 4.14 pN·nm | $10^{-5}$ pN/nm² | 10 pN/nm² |

**Table II** Default parameters for the Langevin chain model in the polymer chain network.

| $l_d$ | $P$ | $L_c$ | $l_0$ | $kT$ | $G$ | $\kappa$ |
|---|---|---|---|---|---|---|
| 24 nm | 0.4 nm | 80 nm | 8 nm | 4.14 pN·nm | $10^{-3}$ pN/nm² | 100 pN/nm² |

**Table III** Default model parameters for porous hydrogels used in the simulation.

| $\mu$ | $\lambda_m$ | $a$ | $t$ | $F_c$ |
|---|---|---|---|---|
| 0.005 MPa | 1.2 | 2 μm | 0.5 μm | 500 pN |